# Combinatorial Sample-and Back-Focal-Plane (BFP) Imaging. Pt. I: Instrument and acquisition parameters affecting BFP images and their analysis

Omer Shavit,[1,2, a] Hervé Suaudeau,[1] Carine Julien,[1,3] Hodaya Klimovsky, [2] Natalia Mañas-Chavernas,[1] Adi Salomon,[2, ID] and Martin Oheim[1,ID,*]

[1] Université Paris Cité, CNRS, Saints Pères Paris Institute for the Neurosciences, F-75006 Paris, France;
[2] Institute of Nanotechnology and Advanced Materials (BINA), Department of Chemistry, Bar-Ilan University, Ramat-Gan, Israel;
[3] Université Paris Saclay, Ecole Normale Supérieure Paris-Saclay, CNRS, PPSM, F-91190 Gif-sur-Yvette, France.

[ID] ORCID id: <u>orcid.org/0000-0001-8139-167X</u> (MO), <u>orcid.org/0000-0002-5643-0478</u> (AS)
[*] <u>martin.oheim@u-paris.fr</u>

ABSTRACT The back-focal plane (BFP) of a high-numerical aperture objective contains the fluorophore radiation pattern, which encodes information about the axial fluorophore position, molecular orientation and the local refractive index of the embedding medium. BFP image acquisition and analysis are common to conoscopy, $k$-space imaging, supercritical-angle fluorescence (SAF) and single-molecule detection, but they are rarely being used in biological fluorescence. This work addresses a critical gap in quantitative microscopy by enabling reliable, real-time BFP imaging under low-light conditions and/or short exposure times, typical of biological experiments. By systematically analyzing how key parameters - such as Bertrand lens position, defocus, pixel size, and binning - affect BFP image quality and SAF/UAF ratios, we provide a robust framework for accurate axial fluorophore localization and near-membrane refractive-index measurements. The described hardware- and software integration allows for multi-dimensional image-series and online quality control, reducing experimental error and enhancing reproducibility. Our contributions lay the foundation for standardized BFP imaging across laboratories, expanding its application to dynamic biological systems, and opening the door to machine learning–based analysis pipelines. Ultimately, this work transforms BFP imaging from an expert-dependent technique into a reproducible and scalable tool for surface-sensitive fluorescence microscopy.

SIGNIFICANCE  Supercritical Angle Fluorescence (SAF) improves sensitivity and resolution in biological microscopy, especially when combined with techniques like TIRF or single-molecule localization super-resolution microscopy, and it has applications in surface sensing and spectroscopy. SAF enhances the surface specificity by acting on the detection side, while TIRF controls excitation near the interface. However, SAF detection relies on digital aperture filtering, meaning that BFP images must be acquired, segmented, and quantified. While sample-plane image quality is well understood, image quality assessment of BFP images is still developing. Our study provides a quantitative analysis of how different experimental parameters affect BFP image information.

---

[1] Part of this work was performed in the framework of OS's PhD thesis, '*Quantitative evanescent-field imaging : Smart surfaces and concurrent sample-plane and back-focal plane image analysis*'. Université Paris Cité (2024), EDPIF (physics), Paris, France.

## INTRODUCTION

Fluorophore radiation patterns carry unique information not captured on conventional sample-plane (SP) images. Back-focal plane (BFP) or, equivalently, 'aperture-', 'pupil-plane' or '*k*-space 'imaging is common in magnetic resonance imaging (MRI), electron microscopy, conoscopy, and some single-molecule detection techniques. In contrast, the acquisition and analysis of BFP images is rarely being used in biological fluorescence. In specialized near-surface fluorescence techniques, supercritical-angle fluorescence (SAF) [1-3] has emerged as an alternative and complement to total internal reflection fluorescence (TIRF) [4-6]. Both TIRF and SAF offer a similar surface selectivity, and they operate on comparable axial ranges of $\sim \lambda/2$ with potentially nanometric precision [7, 8]. TIRF uses evanescent waves to selectively excite near-surface fluorophores, whereas SAF selectively captures emission from surface-proximal fluorophores that radiate into 'forbidden' angles beyond the critical angle.

Like TIRF for excitation, SAF relies on evanescent waves (EWs) but on the emission side [9]. The surface selectivity of SAF arises from the distance-dependent change in the fluorophore radiation pattern near a dielectric interface. When close enough, an emitter's near-field interacts with the interface, allowing evanescent radiation components to become propagating at supercritical angles - a phenomenon akin the leakage of light through the air gap in Newton's air-gap double-prism experiment [10]. Using supercritical light for excitation and emission, respectively, both TIRF and SAF often rely on the use of very high-numerical aperture (NA) objectives [9, 11].

Publications using TIRF still largely outnumber those with SAF[2]. Since the early 2000s, there is a growing body of literature employing SAF for surface spectroscopy [12, 13], imaging molecular orientation [14-16], for axial super-resolution in localization-type super-resolution microscopy [17-22], micro-refractometry [23-25], apertometry of high-NA objectives or solid-immersion lenses [23, 26], nanoplasmonics [27], or for incidence-angle calibration in TIRF [28-31]. We recently used BFP imaging for the axial metrology of fluorescent thin films with nm resolution [31, 32].

Combining TIRF excitation with SAF detection enhances axial sectioning beyond that achieved with TIRF alone [33, 34] but the underlying mechanisms remain unclear, as both techniques are optical reciprocals and - in principle - operate on the same length-scale.

## Imaging fluorophore radiation patterns

TIRF provides optical sectioning by restricting excitation to a shallow layer above the surface, with a penetration depth set (in theory) by the wavelength and refractive indices (see, however, ref. [35]). Axial sectioning by SAF is more complex, as the fluorophore distance but also other factors affect the angular distribution of the emitted fluorescence. The measured radiation pattern also depends on factors like the collection NA, the fluorophore orientation, and polarization effects in the excitation and emission optical paths, respectively (see (8, 9) for review). Also, it is important to realise that back-focal plane (BFP) images gather the *average* emission pattern of *all* fluorophores excited in the field-of-view (FOV) of the objective, with randomly oriented fluorophore distributions producing a centrosymmetric, concentric BFP image. For near-interface fluorophores, SAF corresponds to the thin annular region between the critical angle ($NA_c = n_1$) and the objective's effective NA, $NA_{eff}$. Accurate BFP-image analysis (13) typically via segmentation into SAF and undercritical angle fluorescence (UAF) is essential for SAF microrefractometry (23-25) and axial fluorophore localization (20, 31, 32, 36, 37) and apertometry of high-NA objectives (23, 26).

---

[2] 3,189 (1976-2025) for TIRF *vs.* 68 (1989-2025) for SAF according to *https://pubmed.ncbi.nlm.nih.gov/*

SAF can be acquired on any microscope fitted with a sufficiently high NA-objective ($NA_{eff} \gg n_1$). The radiation pattern is imaged by inserting a Bertrand lens (BL) into the emission optical path. This 'phase telescope' shifts the focus from the SP to the BFP. Unfortunately, today's commercial microscopes lack this option, so that researchers need to design and a BFP collection path on their own. BFP image quality and segmentation into SAF/UAF emission zones depend on the choice of the right optical elements, but also on the BFP image quality and segmentation into SAF/UAF emission zones and other BFP image parameters, such as pixelation, signal-to-noise ratio, and alignment. However, as opposed to SP imaging, no standard protocol or quality metrics exist for BFP images.

With our current work, we want to fill that gap. We here present a hard- and software architecture for combining SP and BFP imaging under low-light conditions typical for biological microscopy. We identify key influences on BFP quality - such as Bertrand lens position, sample defocus, pixel size and binning- and we show how the measured intensities and SAF/UAF ratios (that are directly related to the axial fluorophore position) vary with these parameters. We equally introduce a custom GUI that supports streamlined, multi-parameter acquisition and analysis of both SP and BFP images that is available upon request from the lead author.

## EXPERIMENTAL PROCEDURES

### *TIRF-SAF microscope*

We assembled a custom inverted microscope, combining a scanning TIRF excitation optical path (38) with a motorised emission path that allows rapid toggling between SP and BFP imaging as well as dual-colour emission and multi-spectral imaging. By opting for an open-frame architecture similar to Ref. (33), we have full control over various parameters, Fig. 1a. A detailed part list is found in Table S1 of the Supplementary material and Fig. 1b shows simplified layout of our microscope.

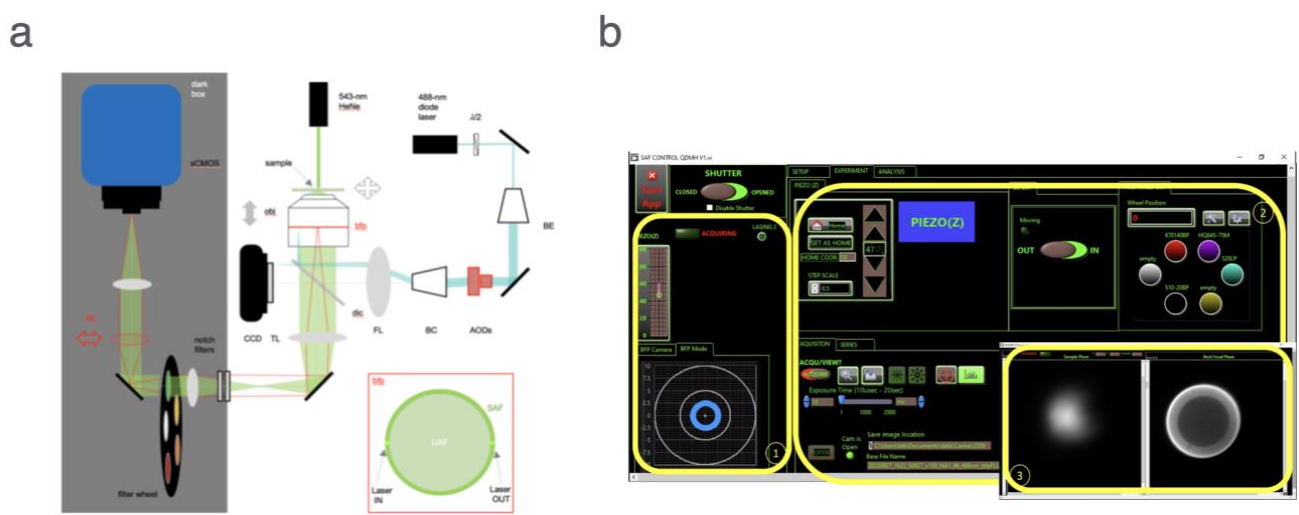

FIGURE 1. *Hardware and Software components for combined BFP and SP imaging.* (a), simplified optical layout of our TIRF-SAF microscope: 488-nm laser; *λ*/2- half-wave plate; AODs- acousto-optic deflectors (beam scanner); BE- beam expander; BC- beam compressor for increasing the scan angle; FL- focusing lens; bfp- backfocal plane; dic- dichroic mirror; CCD- charge-coupled device (BFP camera); BL- motorized Bertrand lens; sCMOS- scientific complementary metal-oxide sensor (main camera); grey shading indicates the elements enclosed in the light-tight detector box sealed off with notch and long-pass filters. A second 543-nm green HeNe gas laser was used for aligning the emission optical path and BL. *Inset*, schematic bfp image, with incoming and outgoing laser spots shown in a TIRF geometry, UAF- under critical angle fluorescence; SAF- supercritical angle fluorescence. (b), A custom graphical user interface (GUI) was programmed in LABVIEW to control the microscope peripherals. Yellow boxes identify different functional areas. From *left* to *right*: (1), instrument status; (2) main acquisition tabs (Setup, Experiment, Analysis), and (3), the result window (3) - which will be generally displayed larger on a second screen, with side-by-side the acquired sample-plane SP (*left*) and BFP images (*right*), which are independently refreshed according to the actual BL position and using their respective binning, exposure-time and contrast settings. A detailed

account of the different functions of these windows along with additional information on the layers hidden on this figure is given in the Supplementary material.

Briefly, the beam of a 488-nm diode laser (Cobolt AB, Solna, Sweden) is attenuated with neutral density filters (ND1.3), expanded ×2.8-fold with a Keplerian telescope, and scanned in $\vartheta'$,$\varphi'$ by a pair of crossed acousto-optic deflectors (AODs, AA-Opto, Orsay, France). The AODs allow for statically positioning the excitation spot at any position within the objective's back pupil, but also for rapid azimuthal beam spinning at fixed polar beam angle (spTIRF) or for classical variable (polar-) angle scans in variable-angle TIRF (VA-TIRF). The scanned beam is re-compressed to double the scan angles otherwise limited by the AODs. While irrelevant for higher magnifications this is important for the ×60 objective that has a larger pupil than the other objectives (see below). An aplanatic *f*=135 mm, 6-element scan lens (Rodagon, Rodenstock, Germany) mounted on an *xyz*-fine precision translation mount focuses the laser beam to a tight spot in the objective BFP. In most experiments, we employed an αPlan-Apochromat ×100/1.46 Oil DIC M27 objective (*f* = 1.65 mm, Carl ZEISS, Oberkochen, Germany). A PlanApo ×60/1.45 Oil TIRF (*f* = 3 mm) or UAPON150X0TIRF ×150/1.45 objective (*f* = 1.2 mm) were used in some experiments (both from Evident, Hamburg, Germany, formerly Olympus Lifesciences). The objective lens is precisely centered on the optical axis by an open-bore *xy* precision-translation mount. Its position along *z* is controlled with a piezo-electric focus drive (P-721, PIFOC, Physik-Instrumente, Germany). A motorized 3-axes stage (HS6, Märzhäuser, Wetzlar, Germany) permits both fine (μm) and long-range (cm) *x,y,z* sample positioning under joystick or computer control. Finally, angular positioning of the sample relative to the microscope's optical axis (i.e., in axes equivalent to the laser scan angle $\vartheta$, $\varphi$) is achieved by a manual open-bore tip-tilt stage (Newport, Irvine, CA). A small monochrome CCD camera (DCC1545, Thorlabs) picks up the transmitted fraction of the excitation light from the main dichroic to provide a real-time BFP image for laser alignment and online visualization of the scan pattern (see GUI).

Fluorescence is collected through the same objective, separated by an ultra-flat (2-mm thick) 491LP dichroic mirror (AHF, Tübingen, Germany) and refocused with a ZEISS tube-lens assembly ($f_{TL}$ = 165 mm). As a consequence, the Olympus objectives have a ≈10%-smaller effective magnification (by a factor 165/180) than manufacturer-specified, reflecting the tube-length mismatch (see Table 1 for details). The combination of a normal-incidence dichroic (HQ490LP, AHF, Tübingen, Germany) and a custom laser-blocking filter (488-561-nm rugate double notch filter, Barr Assoc., Westford, MA) shield the detector box against the totally reflected beam. Depending on the very experiment, fluorescence is further narrowed down through band-pass filters housed in a motorized filter wheel (FW103H/M, Thorlabs, see Supplementary Table S2 for details). Fluorescence is imaged via a relay telescope ($f$ = 110 mm; $f$ = 200 mm) offering an extra ×1.82 magnification onto a water-cooled scientific complementary metal-oxide sensor (sCMOS) detector (PCO edge4.2, Kelheim, Germany) featuring a 4.2 MPx (2048 by 2048 pixels) chip. Pixel size is 6.5 μm. The measured pixel sizes in the SP were 58 nm/px for the ×60, 33 nm/px and 23.8 nm for the ×100 and ×150 objectives, respectively, affording a 4-by-4 pixel-binning for the ×100 and ×150 lenses (and 2-by-2 binning for the ×60)[3] without resolution loss (see below). The entire emission optical path is enclosed in a light-tight box (shaded grey on panel *B*). Supplementary Fig.S1 shows the laser, filter and camera spectra. The Bertrand lens ($f_{BL}$ = 100 mm)

---

[3] *The apparent difference in the effective magnification comes from the fact that Zeiss and Olympus objective are specified with respect to their respective manufacturer tube lenses of $f_{TL}$ = 180 mm and 165 mm, respectively. With our 165-mm tube lens (Zeiss), the two Evident (Olympus) objectives thus have a slightly smaller effective magnification of M' = $f_{TL}/f_{obj}$ = 165 mm/3 mm = ×55 (instead of ×60) and of ×137,5 (instead of ×150), respectively. This impacts on the effective pixel size in the SP, and is important for the on-chip binning to be within the Nyquist limit.*

is mounted on a motorized flipper (MFF101/M, Thorlabs), which allows to toggle between SP- and BFP-image acquisitions with <200-ms dead time. The BL can slide along the optical axis on its dove-tail mount. Depending only on the total magnification of the emission optical path, but not on the very objective used, the BFP image pixel size was 5.15 µm/px in the absence of on-chip binning,

In our free-space optical bench system we use a green HeNe gas laser (543 nm, N-LGP-393 Melles Griot, MicroContrôle Spectra-Physics, Beaune la Rolande, France) to define the optical axis. When injected through the objective this reference beam generates a focused spot in the BFP and in all conjugate aperture planes, which allowed us the axial positioning of the BL, Fig.1c. The BL position differs slightly among the three objectives according to the BFP location inside the objective lens, a value that is difficult to obtain from the manufacturer. The 488-nm laser intensity, laser shutter (LS-3 Uniblitz and VCM-D1 controller, both from Vincent Assoc., Rochester, NY), the AOD scanner, motorized half-wave plate, piezo, BL flipper, filter wheel and the acquisition parameters of both cameras are all controlled through a custom GUI.

### Graphical user interface (GUI)

We developed in a LABVIEW 2020 environment (National Instruments, Austin, TX) a custom GUI to control all peripherals of our setup, see Fig. 2 and Figs. S3-S11 in the Supplemental material online. The GUI was designed to guide the user intuitively through the workflow of combined SP- and BFP-imaging experiments, and it allows for complex, nested multi-parametric acquisitions. The different tabs encourage the user to start systematically from setting up the parameters of each hardware component, to verify the laser beam in the BFP on the second camera, to proceed with a routine alignment-check by focusing on a simple, inexpensive test sample (a µm-thin pyranine layer (39)) and to verify the stage tilt and focus, optimize, if necessary, the BL $z$-position before loading the sample to be studied. Our GUI then invites to independently optimize (and then keep) the image acquisition parameters of both SP and BFP images, which typically have different integration times and binning. A 'live' imaging (streaming) mode is available, as are 'snapshot' acquisitions or automated acquisitions of various image series. We implemented $z$-scans, time-lapse acquisitions, alternate SP- and BFP acquisitions, variable-azimuthal ($\varphi$) or polar ($\vartheta$) angle scans for VA-TIRF protocols. Likewise, emission spectral scans permit changing between pre-defined band-pass filters for a coarse spectral fingerprinting and linear unmixing (40). All acquisition protocols can be 'nested', e.g., spectral scans be acquired in a time-resolved manner.

Our GUI also caters for consistent data management: generic filename roots are set up and are automatically appended with suffixes indicating the excitation wavelength, filters in place, the acquisition mode (EPI or TIRF, azimuthal and polar beam angle on the excitation side; SP or BFP on the emission side). Identical file roots are appended with trailing numbers in the case of multi-plane acquisitions and the acquired images are automatically sorted into subfolders according to the respective acquisition protocols for facilitating later access. Our GUI thereby facilitates data organization and handling, it sets up a uniform file-name format and supports a multi-parametric image acquisition, an important feature for image 'housekeeping', particularly for nested multi-plane SP and BFP images stacks that generate large data sets in a single experiment. The GUI also contains functionalities for the online diagnostics as well as indicators of the system status, a real-time view of stage position (similar to an 'artificial horizon'), and a real-time view of the excitation-light distribution in the BFP, which allows monitoring beam scans in real time (see SI).

Unless otherwise stated, images were scaled to a black-to-white (grey value) look-up table (LUT) between min-max counts in the image after subtraction of the average background. Unless explicitly otherwise stated, image contrast is linear ($\gamma$=1). In low-light cases we subtracted an average ($n$=10 frames) dark image taken with the same acquisition parameters but the laser shutter closed.

*Online image analysis*

We provide an on-line image analysis tool, coded in MATLAB (The Mathworks, Natick, MA) and integrated into the LABVIEW environment (Fig. S12). Its functions include an area-detection algorithm that first delineates three BFP-image zones: background, super- (SAF) and under critical emission (UAF) components, respectively, by dual-threshold segmentation (Fig. S13). It then calculates the average background intensity, and, respectively, the cumulative intensities (after background subtraction) of the SAF and UAF image zones, their sum, $S = I_{SAF} + I_{UAR}$ (total fluorescence), as well as their ratio, $R = I_{SAF}/I_{UAR}$ (see results).

The algorithm also provides three figures of merit, the *circularity CI* ($4\pi$·area/perimeter$^2$) (41), the *ovality OV* = 2[*a-b*]/[*a+b*] (where *a* and *b* are the lengths of the major and minor axes of an ellipse fitted with the thresholded image, respectively. This dimensionless quantity is between 0 and 1). We also estimate the *concentricity CO* (i.e., the absolute difference in position of the centeres of the outer and inner circle) for the UAF and SAF zones for routine quality assessment. Together, these parameters provide a metric for verifying the alignment of the elements of the excitation and emission optical paths.

In addition, a first estimate is generated of the radius that separates UAF from SAF emission components (i.e., the RI $n_1$ of the fluorophore-embedding medium) using the calculated equivalent (constant) radius of the UAF area and the *à priori* knowledge about the independently measured (26) effective NA of our objective, $NA_{eff}$ =1.461±0.009 for the ZEISS lens (in close agreement with ref. (23) where we employed the same objective) and 1.427±0.003 and 1.433± 0.005 and for the ×60 (Fig. S14) and ×150 Olympus lenses, respectively. We note that both Olympus lenses had effective NAs that were considerably smaller their specified values.

*Offline image analysis*

A stand-alone algorithm allows for image segmentation and fitting. SP and BFP images are subtracted with their respective (identical exposure and binning) dark images. BFP images are segmented into SAF and UAF regions in a multi-step process: we first find the center of the BFP pattern and the number of pixels corresponding to the limiting radius NA$_{eff}$. To this end, we binarize the image thresholding first for background and an object recognition algorithm implemented in MATLAB finds the disk that best fills the SAF+UAF area. (The $NA_{eff}$ measurements were generated in this manner). Next, we emit a first guess for the pixel radius at the critical angle of the radiation pattern ($r_c$) by imposing a refractive index ($RI$), $r_c = r_{NA}RI/NA_{eff}$. Integration is performed over all pixels within the bounds $r = (0, r_c)$ for UAF and $r = (r_c, r_{NA})$ and for SAF. Finally, error bars are generated by allowing for a positive and negative deviation (d$RI$) according to $r_c^{(\pm)} = r_{NA}(RI \pm dRI)/NA_{eff}$. This procedure is repeated with different $RI$ estimates. Together, the found center coordinates, $r_{NA}$ and $r_c$ enabled us to reliably segment the image into three areas: background, SAF and UAF. Again, the average background in the image zone outside the circles is subtracted and the intensities of the UAF and SAF areas, respectively integrated. The obtained intensities are finally used to calculate the total fluorescence $S = I_{SAF}+I_{UAR}$ and the SAF/UAF ratio, $R = I_{SAF}/I_{UAR}$, where I is the respective integral over SAF and UAF areas. Repeating this integration for each dRI provided an estimate of the dependence of SAF/UAF ratio on the accuracy of the RI estimate.

We then generated an estimate of the azimuthal emission (an-)isotropy. For isotropic emitters, the BFP image should display radial symmetry. This symmetry is broken if fluorophores are non-randomly oriented, in the case of a detection bias or for effects resulting from excitation or emission polarization (6). Other sources of asymmetry are misalignment, tip-tilt of the substrate with respect to the optical axis, vignetting, etc. To quantify this azimuthal asymmetry, we measure the intensity along a user-specified circular band having a width $r_2$-$r_1$. To this end, a circular profile is generated at $r = (r_2+r_1)/2$ , where $0 < r < r_{NA}$ and

$$I(\phi) = \int_{r_1}^{r_2} I(r,\phi)dr \qquad (eq.1)$$

calculated and represented as a function of azimuth $\phi$ (polar plot).

## Calibration samples

*Test targets, rulers and reticles.* We employed a transparent µm-slide (R1L3S2P, Thorlabs) for the calibration of SP pixel sizes, a small mm-sized transparent glass ruler (Glass Measuring Ruler, Pepler Optics, Knutsford, UK) for BFP pixels, and a flat mirror - at the position of the BFP - for the calibration of command voltages of the AODs in terms of BFP pixels and for the Thorlabs BFP camera pixels.

*Textmarker film.* Pyranine (CAS: 6358-69-6, from a yellow highlighter pen, StabiloBoss, Schwan-Stabilo, Heroldsheim, Germany) was applied in a single swipe on a clean #1 BK-7 glass coverslip (Menzel-Gläser, Braunschweig, Germany) and left in the dark to dry overnight (39). In this way, a 20-µl drop of either water (or a glycerol/water mix) could be deposited on top without dissolving the dye. We used this inexpensive test sample prior to each experiment as an reference standard for quality control (QC), with an EM520LP long-pass emission filter.

*Fluorescent microspheres.* 93-nm diameter yellow-green FluroSpheres™ Carboxylate-Modified Microspheres (ThermoFisher) were diluted to 1% volume in MeOH by taking 1 µl of the commercial solution and adding 99 µl of MeOH. A 20-µl drop was deposited on a clean BK-7 glass coverslip and left to dry overnight in a dark environment prior to use.

*Nanolayered sandwiches.* Ultra-thin sandwiches alternating spacer and dye layers were deposited on cleaned #1 BK-7 glass substrates (Marienfeld) as described elsewhere (31). Briefly, non-fluorescent, transparent My-133-MC polymer spacer layers (MY Polymers, Ltd., Ness Ziona, Israel) were spin-coated. A thin 5,10,15,20-tetrakis(4-sulfonatophenyl) porphyrin ($H_6$TPPS) J-aggregate dye layer was drop-cast and immediately blown off with a $N_2$-jet (42). Finally, a µm-thick My-133-MC capping layer was applied to seal off the dye-layer in a homogenous refractive-index environment. Samples were thoroughly characterized by a combination of profilometry, AFM, electron microscopy and ellipsometry prior to use (not shown).

In some experiments, we used bright, thin fluorescent monolayers created from proprietary nano-bead emitters (NBEs) (43). NBE layers were produced by a modified layer-by-layer (LBL) method. Briefly, a primer layer of PDDA (poly(diallyldimethylammonium chloride), Sigma Aldrich #544981) was applied to a corona-treated surface to enhance adhesion. The NBEs themselves consist of negatively charged hydrogel nanoparticles (23±2 nm diameter, ζ potential of -9.13 mV , measured by DLS) that were labeled with ATTO488 dye via covalent amide bonds. Once dried, the resulting dye films form smooth, uniform and brightly fluorescent thin films of 6- to 10-nm thickness. A detailed characterization for these NBEs is published elsewhere (44).

# RESULTS

## Different axial fluorophore profiles produce distinct emission patterns

Epifluorescence excitation (EPI) of an aqueous solution of pyranine (the yellow dye in fluorescent highlighter pens (39)) produced a near-uniform fluorescence distribution in the back-focal plane (BFP) of a NA-1.46 objective, Fig. 3a, *top*. The corresponding equatorial intensity profile, Fig. 3b, (*red*) is not a perfect flat-top function as it is affected in an emission angle-dependent manner by the Fresnel coefficient of transmission and by the spatial non-uniformity of the objective's collection efficiency. As a consequence of the bright volume fluorescence upon EPI excitation we note some 'leak' of signal even beyond the nominal collection aperture (*red* arrow), a phenomenon known from the non-imaging collection of fluorescence photons scattered inside the objective, (45). Upon closer inspection, we can note a marked intensity drop at an NA slightly *below* the limiting $NA_{eff}$ of the objective, Fig. 3b, *red* arrowhead. The drop occurs at NA = 1.33, at the NA corresponding to the refractive index (RI) of water (*dashed line*) and marking the emission critical angle $\vartheta_c = \mathrm{asin}(n_1/n_2)$ $\approx 62°$ of water. Here, $n_2 = 1.5214$ is the substrate RI of Schott BK-7 glass of the coverslip used (46).

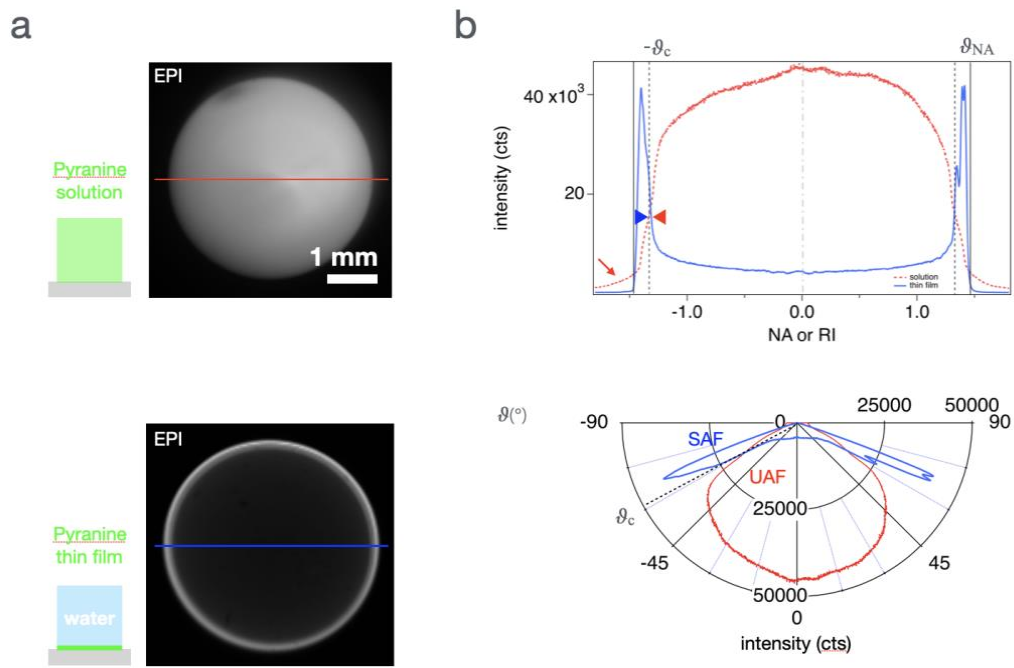

FIGURE 3. *BFP images and fluorophore radiation patterns for different axial fluorophore distributions upon high-NA detection.* (a), BFP images recorded, respectively, upon epifluorescence (EPI) excitation of a pyranine drop (*top*) and a μm-thin film of dried pyranine, topped with water (*bottom*). Scale bar, 1 mm. Objective x100/NA1.46 Plan-Apochromat. (b), *top*, radial intensity profiles along the red and blue cross-sections on the images shown in panel (a) reveal markedly different emission patterns for bulk and surface-proximal fluorophore distributions. *Bottom,* polar plot representation of the same data. Note the pronounced supercritical lobes for the substrate-proximal fluorophores (*blue*),

whereas most fluorescence is emitted into undercritical angles for the bulk pyranine solution (*red*). Dashed line identifies the calculated emission critical angle, $\vartheta_c$ for the BK-7 borosilicate glass/water interface, indicating the transition between SAF and UAF. Colors as on (a). For generating the polar plots, we assumed that the objective fulfilled Abbe's sine condition.

As expected for molecular dipoles mostly located at distances $z \gg \lambda$ above the substrate, their radiation pattern mainly falls within a disk circumscribed by NAs $< n_1$. We nevertheless see some supercritical emission but at a much fainter intensity. The origin of this emission component becomes clear when turning instead to another sample having a very different axial fluorophore distribution, but the same RI, a $\approx$1-µm thin dried pyranine film on glass, topped with water. The radiation pattern was now almost opposite, Fig. 3a (*bottom*), and dominated by a concentric bright ring *above* $\vartheta_c$ and a lower-intensity disk at undercritcal angles. The corresponding fluorescence intensity line profile reveals sharp intensity peaks at angles 'forbidden' for propagating light, Fig. 3b (*blue*). As before, the intensity transition occurs at $n_1 = 1.33$, but the dominant emission was supercritical (*blue arrowhead*). Thus, unlike for the dye in solution, the emission from a thin, near-interface fluorophore film is highly directional and predominantly emitted as supercritical angle fluorescence (SAF). The polar-plot representation of the same data clearly reveals the very different radiation patterns for fluorophores very close to (*blue*) and remotely (*red*) from the substrate, respectively, Fig. 3b (*bottom*). As expected from theory (9), depending on the axial fluorophore profile, we observe very different fluorescence emission patterns.

**A quality metrics for BFP images**

We routinely started our experiments by imaging the pyranine film used in fig.1a (*bottom*) for verifying alignment, uniformity and signal. In some cases, we noted some asymmetry, offsets or imperfections on the resulting BFP images. Such defaults will affect any BFP-derived measurements, for example of the effective NA$_{eff}$ of the objective (26), of the local refractive index (RI) $n_1$(23), or the dipole orientation (15). To better study the impact of flaws on both SAF and UAF emission components, we 'magnified' the SAF ring by switching to a dye/air rather than a dye/water interface, Fig. 4a (*left*). Colored lines on the BFP image identify four radial intensity profiles at different azimuths. While the image looks uniform and isotropic at first sight, the cross-sections at $\phi = 0°$, 45° and 135° show some differences, particularly at very high emission angles (NAs), Fig. 4a (*right*). These discrepancies might be due to, (*i*), vignetting along the emission optical path; (*ii*), coverslip tip-tilt; (*iii*), residual fluorophore anisotropy; (*iv*), a markedly inhomogeneous spatial fluorophore distribution across the field-of-view (FOV); (*v*), an uneven illumination or collection efficiency across the objective pupil; (*vi*), a non-uniform background, or a combination of these.

Whereas we have a metrics for the quality of SP images based on the objective point-spread function, Nyquist's sampling theorem, field homogeneity and signal-to-noise considerations of the detector (see, e.g., (47) for recent efforts to implement a quality metrics), no such standards exist for fluorescence BFP images, and their quality assessment (QA) is still in its infancy. Giving complementary information along the azimuth, Fig. 4b shows the circumferential intensity profile, which reveals emission anisotropy, if present. For the BFP image shown, the polar plot had a coefficient of variation (CV) of <3%, indicating that fluorophores in the pyranine layer had no preferential orientation and the excitation and emission were largely unbiassed. Some abrupt peaks and dips disrupted the azimuthal intensity profile, likely due to specs an dust on intermediate optical elements in or close to an equivalent BFP or as a result of micro-bubbles in the immersion oil.

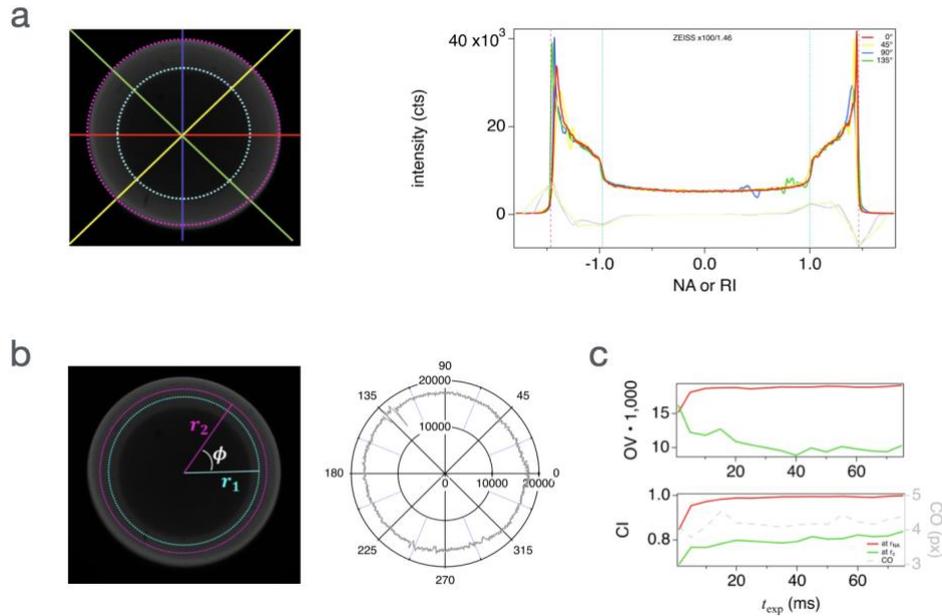

FIGURE. 4. *Quality assessment of BFP images: radial and azimuthal intensity profiles, and figures of merit*. (a), *left*, BFP image of a pyranine on BK-7/air interface with colored lines indicating radial cross-sections at different azimuthal angles, $\phi$ =0°, 45° and 135°, respectively. Ovality (OV) and circularity (*CI*) of the BFP image shown were 0.001 and 0.89, respectively, indicating a high-quality BFP image. *Right*, intensity cross-sections, after background subtraction and scaling of the BFP radius to $\pm$NA. Colour code as before, pale traces show 2nd derivatives that were used to find the points of steepest slope. Dashed purple and cyan lines identify, respectively, the extrema of the 2nd derivative that corresponds to the measured effective NA (1.465) and refractive index (RI) of air (1.002), respectively. (b), *left*, emission anisotropy. An annular band was defined between the turquoise and purple lines ($r_1 \leq r \leq r_2$) and the mean intensity represented as a polar plot, *right*. Tiny air bubbles in the immersion oil or dust and other imperfections on intermediate optical elements show up as kinks on the azimuthal intensity profile. (c), plot the evolution of circularity *CI*, ovality *OV* and concentricity *CO* with BFP image exposure time for the same sample as before. Red and green curves correspond to zones below the critical angle NA$_c$ (in the UAF zone) and close to the limiting NA of the objective (SAF), respectively.

We next acquired BFP images of the pyranine dye/air interface (dye at $z$ = 0) at different exposure times. For each BFP image, we plotted the ovality *OV*, circularity *CI*, and concentricity *CO* (see Methods for the definition of these parameters) as figures of merit, Fig. 4c. The outer pupil and the SAF/UAF transition displayed ovalities <1% and <2%, respectively), indicating that the major and minor axes of the ellipse were virtually identical, Fig. 4c, *top*. Also, the shape of the segmented area barely deviated from a perfect circle[4], as *CI* values were close to unity for the outer pupil and of the order of 0.8 for the SAF/UAF transition, Fig. 4c, *bottom*. The systematically lower value for the inner circle results from the smaller intensity difference and lesser contrast of the SAF/UAF transition compared to the NA/background transition (see below). However, the concentricity, i.e., the center-to-center offset of the inner and outer equivalent circles fitted with the full aperture and under critical aperture, respectively, was typically of the order of 4 pixels, i.e., less than 0.5% of the pupil diameter, indicating the good alignment of the emission optical path. Above 25-ms exposure time, these measurements converged with little spread ($CI_{NA} = 0.98 \pm 0.04$, $CI_c = 0.79 \pm 0.03$, $OV_{NA} = 0.019 \pm 0.001$, $OV_c = 0.011 \pm 0.001$, $\Delta r = (4.2 \pm 0.2)$ px, respectively) and can thus be considered as robust measurements.

Thus, while real-world BFP images differ from immaculate simulations (see, e.g., ref.(9)) due to imperfections in the dye layer, small air bubbles in the immersion oil, dust on intermediate optical elements and distortions in the emission path, they still are fairly symmetric. Yet, even for a well-aligned emission path, they still exhibited some minor non-circularity, non-concentricity, and intensity anisotropy. The parameters defined in this section offer the first comprehensive quality metrics for BFP images. In our experience, the main challenges do not come from misalignment or sample imperfections but often result from too dim images, which produce poor *CI* and *OV* values, complicating segmentation and further analysis (see below).

### Automated segmenting of BFP-images can be tricky for thick samples

We developed a two-step thresholding and fitting procedure to automatically segment BFP images into background (B), supercritical (SAF), and undercritical (UAF) fluorescence zones, respectively, Fig. 5a. This segmentation is essential for intensity-based analyses. We employed MATLAB's histogram-based intensity thresholding algorithm (48) to separate background from signal, ('threshold 1'). An ellipse was then fitted to the resulting binary mask to extract the center and axes. From the ellipse, we generated an equivalent-area circle that defined $r_{NA}$ (or, equivalently, NA*eff*). In a second step, a lower threshold ('threshold 2') was applied to segment SAF from UAF, yielding an annular binary mask and the critical radius, $r_c$ (NA_c).

By initially not enforcing circular symmetry, our method accounts for slight sample tilt or inhomogeneous fluorophore distribution across the field-of-view (FOV). This approach worked in most cases, but failed to segment the BFP image of the EPI-excited pyranine drop (*c.f.* Fig.3a, *top*), Fig. 5b, *top*. In such cases, we imposed a 'critical circle' from the known RI of water ($n_1$=1.33) and iterated this procedure in tiny steps allowing a $\pm 0.05$ RI uncertainty. Iterative fitting revealed a peak in the derivative of the SAF/UAF ratio at the correct transition radius (Fig. S16). The resulting SAF/UAF segmentation was complementary to that of a thin dye film under water. We conclude that intensity-based segmentation without constraints can fail for bright, diffuse emitters lacking sharp

---

[4] Unlike the ellipticity, the circularity compares how much the thresholded free form deviates from a perfect circle, imagine, e.g., a flower- or star-shaped region having the same area as the (smaller) circle.

BFP features. In contrast, our dual-threshold method performed reliably for localized fluorophore distributions and under TIRF excitation-conditions typical in near-surface fluorescence. As expected, TIRF facilitates segmentation by suppressing bulk emission compared to EPI.

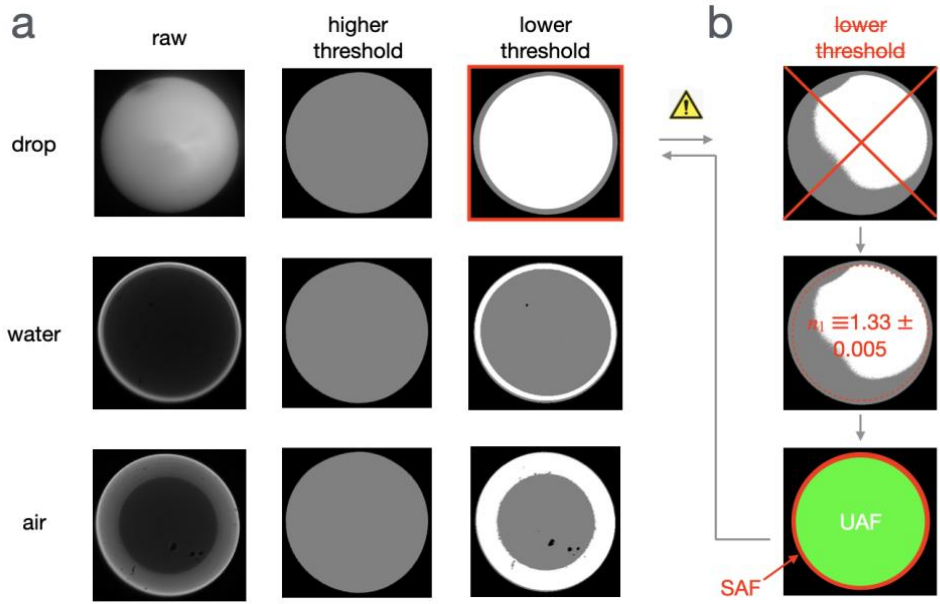

FIGURE. 5 *Histogram-based automatic segmentation of pyranine BFP images into background, SAF and UAF emission zones.* (a), *left*, BFP fluorescence images of a pyranine drop on a BK-7 substrate (*top*), a pyranine film topped with water on BK-7 (*middle*) and an air/pyranine BK-7 interface (bottom) interface, respectively. x100/NA1.46 objective and 520LP emission filter. *Middle*, identical results are obtained for all input images when segmenting background (*black*) from signal (*grey*). *Right*, results of the second, lower-threshold, segmentation to distinguish SAF and UAF. While the *middle* and *bottom* images are well segmented, the aqueous pyranine solution  in (*b*), corresponding to fig.3a (top) is problematic and automatic segmentation fails. (b), as a consequence, we impose a first guess (RI of water, 1.33) and then varied the radius about this value to obtain a consistent segmentation. (See main text).

We next integrated on the images shown in Fig. 5a the fluorescence over the segmented zones and we calculated their sum $S = I_{\text{SAF}} + I_{\text{UAF}}$ and ratio $R = I_{\text{SAF}}/I_{\text{UAF}}$, Table 1. Using a nanometric axial ruler and the same objective, we have previously shown that $R$ is directly related to the fluorophore height above the interface (31). All images displayed low ovality and high circularity, with systematically higher values for the outer pupil ($\text{NA}_{eff}$) than for the inner (SAF/UAF) transition. Finally, the small *CO* offset of the order of 2 to 4 pixels is indicative of a good alignment of the collection optical path and the absence of stage tip-tilt with respect to the optical axis.

Table 1

*BFP image quantification and quality metrics*

| BFP image | R | S (kcts) | $OV_{NA}$ | $OV_c$ | $CI_{NA}$ | $CI_c$ | CO (px) | Comment |
|---|---|---|---|---|---|---|---|---|
| Pyranine drop | 0.12 | 46,175 | 0.0065 | 0.0396 | 0.91 | 0.76 | - [a] | segmented by imposing $n_1$ |
| film/water | 0.80 | 65,017 | 0.0235 | 0.0285 | 0.99 | 0.99 | 2.8 | |
| film/air | 3.09 | 85,1287 | 0.0189 | 0.0093 | 0.99 | 0.79 | 4.1 | |

[a] *UAF disk is centered on outer pupil.*

Taken together, the here defined set of parameters qualifies BFP images in terms of asymmetry and con-centricity and it provides a quality assessment and checkpoint prior to engaging on BFP image analysis and quantifying SAF and UAF emission intensities.

## Bertrand lens defocus has little impact on BFP image segmentation

We mounted the BL on a motorized flipper to rapidly alternate between BFP- and SP-image acquisitions. The BL is an auxiliary lens in an intermediate image plane. It images the BFP of the objective onto the main camera. BFP positions are typically inside the objective and they differ among objective lenses, even of the same supplier. We hence decided to experimentally determine the 'best focus' position of the BL for each objective. To this end, we injected a green HeNe laser beam from above into the objective (see fig. 1c) and we moved the BL in its dovetail mount until we observed the smallest spot on the BFP image.

How sensitive are the resulting BFP images to BL defocus? Imaging a pyranine film and focusing the SP image first, we switched to BFP imaging and introduced systematic positive and negative defocus to the BL, Fig. 6a *(top)*. On the collected BFP images we measured radial intensity profiles as before. Plotting the normalized fluorescence $F/F_{max}$ of each profile, Fig. 6a, we noted two effects of wrong BL position, (*i*), a defocus-dependent shift of the SAF peak lobe of the radiation pattern (Fig. 6a, *inset* a1) and a reduction in slope at the limiting $NA_{eff}$ (Fig. 6a, *inset* a2). Representing the measured peak positions and slopes, respectively, as a function of defocus indicated that both parameters peaked at the best-focus position of the BL, Fig. 6b. Conversely, while the calculated SAF/UAF intensity ratio varied only slightly (<3% for all practical amounts of defocus) it showed a monotonous rise with BL position, Fig. 6c, similar to moving a magnifying glass when displacing the BL relative to the BFP.

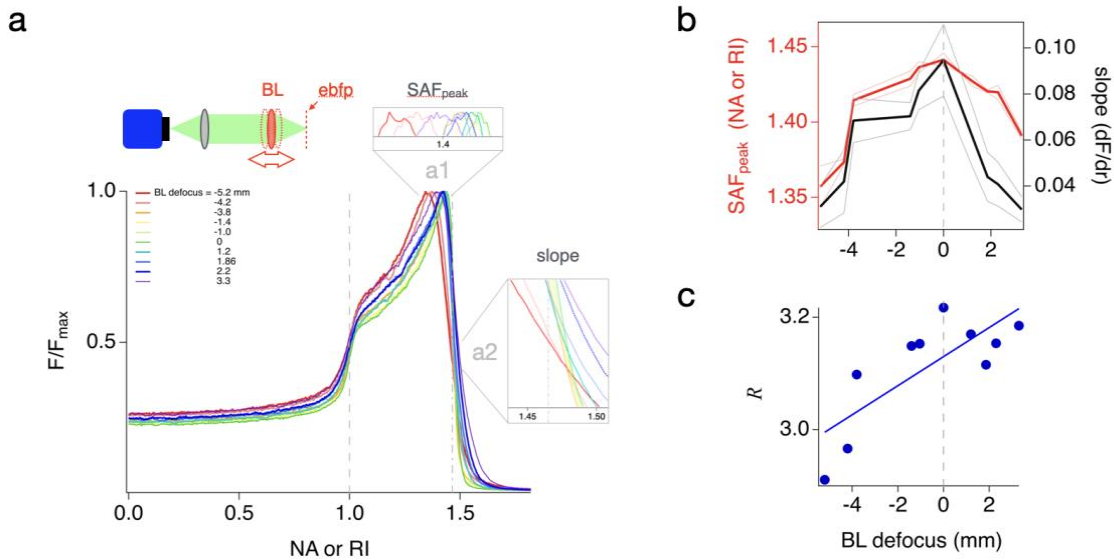

FIGURE 6. *Effect of BL defocus on SP and BFP images.* (a), *top*, experimental paradigm: the Bertrand (BL) lens was displaced along the optical axis and the amount of defocus measured with calibers. *Bottom*, azimuthally averaged radial intensity profiles for different amounts of BL defocus are shown color coded (green corresponds to best focus, red and blue hues to positive and negative defocus, respectively). *Insets* a1 and a2 zoom in on the displacement of the peak of the radiation pattern and change in slope near the limiting $NA_{eff}$ of the objective lens, respectively. (b), measured SAF peak position (in NA units, *red*) and steepest slope near the limiting $NA_{eff}$ (*black*), respectively, as a function of BL defocus. Grey traces are individual measurements, black trace is ensemble average. (c), the SAF/UAF intensity ratio $R$ varies linearly (and little) with defocus. Thus, as expected, moving of the BL changes the magnification of the phase telescope, equivalent to a zoom lens.

Taken together, the alignment of the 'phase telescope' is fairly straightforward, even in the absence of manufacturer information on the precise BFP position. Finding the best focus position of the BL is simple by looking at the slope or peak position of the radiation pattern. BFP image parameters only slightly depend on BL defocus.

**BFP images are more robust against defocus than SP images**

Once the BL fixed in its 'best focus' position we noted that the observed radiation patterns were fairly robust against sample defocus. Indeed, BFP images were much less sensitive to moving the objective lens with the piezo than the corresponding SP images. In practice, it was often easier to first find and focus on the sample by using BFP images, and then adjust the fine focus only after switching to SP image acquisition. To quantify and understand this phenomenon, we prepared a coverslip sparsely sprinkled with fluorescent point sources (93-nm diameter yellow-green emitting beads) and acquired z-stacks of SP and BFP images, respectively. We extracted from SP image stacks of $n = 20$ single beads their lateral and axial the intensity profiles, respectively, Fig. 7a. Upon epifluorescence (EPI) excitation ($\vartheta, \varphi = 0,0$) we measured half-widths (FWHM = $2[2\ln(2)]^{\frac{1}{2}}\sigma$) of 298 nm in $x$ and 680 nm in *z*, respectively (*red* curves show a fit with the data). These values are, respectively, 16% and 45% larger than the calculated lateral resolution (257 nm) and nominal depth-of-field (468 nm) of the ZEISS NA1.46 objective.

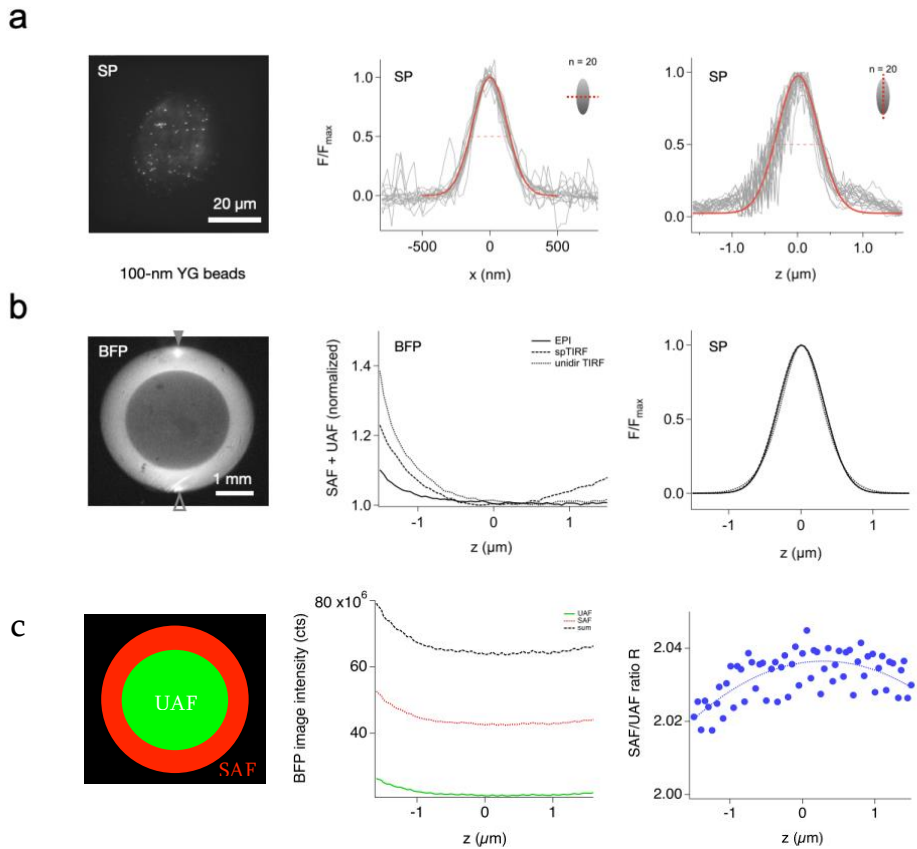

FIGURE 7. *Effect of sample defocus on SP and BFP images.* (a), *left*, example in-focus sample-plane (SP) image of 93-nm diameter yellow-green emitting beads dried on a coverslip in air. *x*- and *z*-intensity profiles were measured from individual beads, (grey traces) and the average lateral, *middle*, and axial, *right*, size of the microscopes effective point-spread function quantified as the FWHM of a Gaussian distribution fitted with the measured intensity (red). Note the slight asymmetry resulting from focusing above and into the coverslip on the top right panel. (b), *left*, BFP image of the same sample as in (a), for unidirectional evanescent wave excitation. The incoming (filled arrowhead) and reflected laser beam (open) are seen in the periphery of the objective pupil. The normalized ($F/F_0$) total fluorescence measured in the BFP varied little upon defocus (*middle*) compared to the corresponding SP images, *right*. (c), *left*, BFP image segmentation and evolution of the SAF (red), UAF (green) and summed intensities (black) with defocus (*middle*). The calculated SAF/UAF ratio, *right*, varied only slightly, confirming the observed relative robust-ness of BFP images with defocus.

Analysis of the corresponding *z*-stack of BFP images revealed a lesser axial dependence of the BFP image intensity (integral of UAF + SAF), confirming their relative robustness against defocus, Fig. 7b, *middle*. As expected, the *z*-dependent obstruction of the objective's pupil depended on the precise distribution of excitation light, with unidirectional and spinning TIRF excitation being more sensitive than EPI. No such effect is apparent on the corresponding SP images where all three curves overlapped, *right* panel. We finally segmented the BFP images of the acquired stack as before and plotted $I_{UAF}$, $I_{SAF}$, their sum $S$ and ratio $R$ as a function of defocus, Fig. 7c. Substantiating our earlier observation, $R$ varied little with defocus and displayed a bell-shaped dependency on d$z$ with a maximum at best focus of the objective. Relative deviations for realistic amounts of defocus were <1% of peak value, little more than the measurement accuracy of $R$ itself, Fig. 7c, *right*. This is plausible, as the BFP contains the decomposition into plane waves of what is happening in the SP. Defocus introduces a parabolic phase shift to the plane-wave decom-position, but this does not change the amplitudes. We would expect that as long as the defocus is not obstructing the entry pupil of the objective, there should be no spatial filtering of the beam. This also explains why TIRF excitation (spinning or unidirectional) was more sensitive to defocus than EPI (fig.7b, *middle*), as the excitation light passes through the extreme periphery of the objective pupil.

We conclude that BFP images are fairly insensitive against defocus and handy for grossly finding the focal plane. Conversely, they are ill-suited for precise focusing. It is thus preferably to image the SP before the acquisition of quantitative BFP image series. Such combinatorial SP and BFP imaging is supported by our GUI. On the other hand, BFP images and BFP-image derived parameters are comparably tolerant to focal drift and slight defocus, which is an good news for long imaging sessions, as often used for capturing biological dynamics and for experiments in which no active focus drive feedback is being used.

**On-chip binning improves BFP image segmentation**

In addition to the accuracy and precision of finding the critical and limiting radii, respectively, subsequent intensity measurements depend on the signal-to-noise ratio, which, in turn will depend on the BFP-image exposure and on the pixel size. In a given experiment, a compromise thus must be found between having small enough pixels to precisely delineate SAF and UAF regions on the one hand and collecting enough photons per pixel on the other hand. Without binning, our BFP pixel size was 5.15 µm/px, meaning that the BFP image covered only about ¼ of the large-format sCMOS sensor, so that - technically - up to 4-fold on-chip binning can be used for maximizing the signal.

Combining BFP images of the same sample taken at different exposure times or pixel bin, we measured radial intensity profiles, Fig. 8a, which differed in their intensity and noise, but - by and large - these overlaid well when normalized to peak. Indeed, the integrated SAF and UAF intensities ($I_{SAF}$ and $I_{UAF}$, respectively) scaled linearly with exposure time Fig. 8b (*red and green*), independently of the binning used. Their ratio, $R$, was constant once above the noise limit, Fig. 8b (*blue*). Thus, binning of BFP image pixel is an effective means to increase signal and temporal resolution when short exposure times are mandatory or the signal is low. For all binning and exposure times studied,

SAF/UAF ratios were a robust measure and data could be combined even from experiments with different binning or exposure times. Yet, binning changes the spatial sampling frequency, which brings us to the somewhat puzzling question of what BFP pixel size we should use (see below).

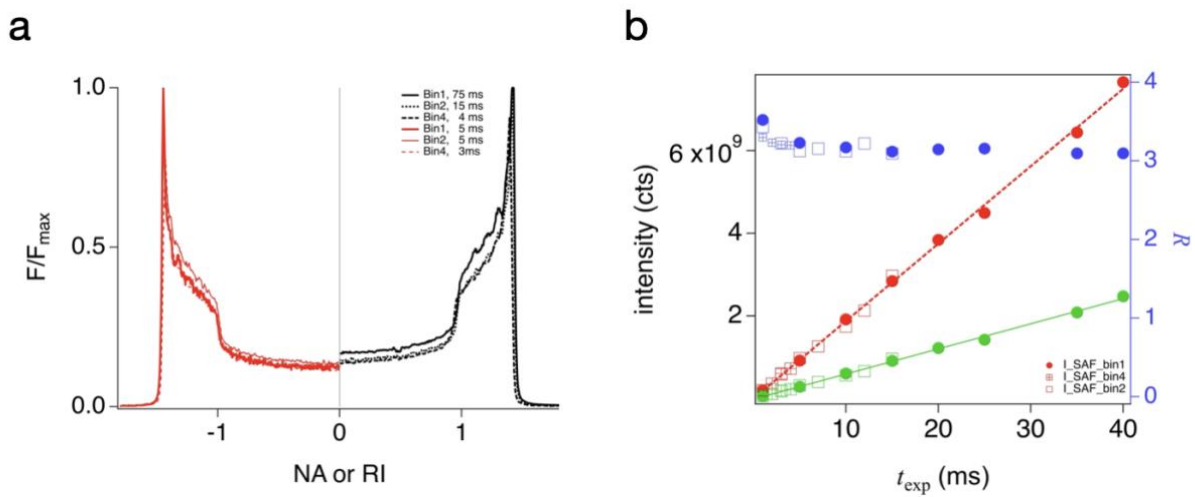

FIGURE 8 *SAF/UAF ratios are independent of pixel binning and exposure time.* (a), radial intensity line profiles of half-BFP images upon on-chip binning 1×1, 2×2, 4×4. *Left*, (black) intensity profiles for 3- and 5-ms exposure time. *Right*, same for 4-, 15- and 75-ms exposure just before reaching pixel full-well capacity, *red*. Profiles are normalized to peak intensity ($F/F_{max}$). (b), plot of the cumulative intensities, obtained from the same images that were used for generating the cross-sections in (a), integrated over SAF (*red*) and UAF zones (*green*), respectively, and their ratio $R = I_{SAF}/I_{UAF}$ (*blue*). BFP image data from different exposure times and binning were pooled. The respective background intensities were subtracted. Note the overall linearity and the constant $R$ above 10-ms exposure.

## Experimental radiation patterns depend on the objective used

We imaged the same pyranine fluorescent-film sample on the same microscope but with three different objectives. The objectives had similar NAs in the range of 1.45-1.46 but magnifications from ×60 to ×150, Fig. 9a, *top* and Table 2. These samples are homogeneous on a length-scale much larger than the (illuminated) fields-of-view (FOV).

We first estimated the FOV of the three objectives from SP images of a thin pyranine film by a bidirectional cross-sectional analysis to be about 50 μm by 50 μm, (34 μm)$^2$ and (22 μm)$^2$ for the ×60, ×100 and ×150 ×150 objective, respectively, the exact value and shape depending on the type of illumination (EPI, unidirectional or spinning-TIRF, see Fig. S17 for details). 'Illuminated' FOVs are considerably smaller than the nominal FOVs due to the limited focusing and finite size of the exciting laser beam in the BFP. Fig. 9a, *bottom*, shows the resulting BFP images. As a result of the different magnifications, BFP images have different formats ($\propto 2 \cdot NA_{eff} \cdot f_{obj}$). To our surprise, despite their similar NAs, the measured equatorial intensity cross-sections differed considerably among the objectives, Fig. 9b. The ZEISS objective not only captured higher SAF intensities (which is expected from its higher transmission compared to the other lenses, see fig. S18), but it was the only one to resolve the fine, high-intensity peaks at very high collection angles. These features were completely absent from the BFP images and line profiles acquired with the two OLYMPUS lenses. This is plausible, as our two OLYMPUS lenses had a lower measured (*effective*) $NA_{eff}$ (both around 1.43 vs. 1.465) than the ZEISS objective (1.465). Different degrees of optical corrections could exacerbate these differences. As a consequence, for the same sample the calculated SAF/UAF ratio was $R = 3.08 \pm 0.03$ for the ZEISS objective, almost double than that obtained with the ×60 and ×150 OLYMPUS lenses ($R$=1.61±0.03 and 1.44±0.02, respectively). Thus, the *experimental* radiation patterns and any measurements derived thereof depend on the objective used.

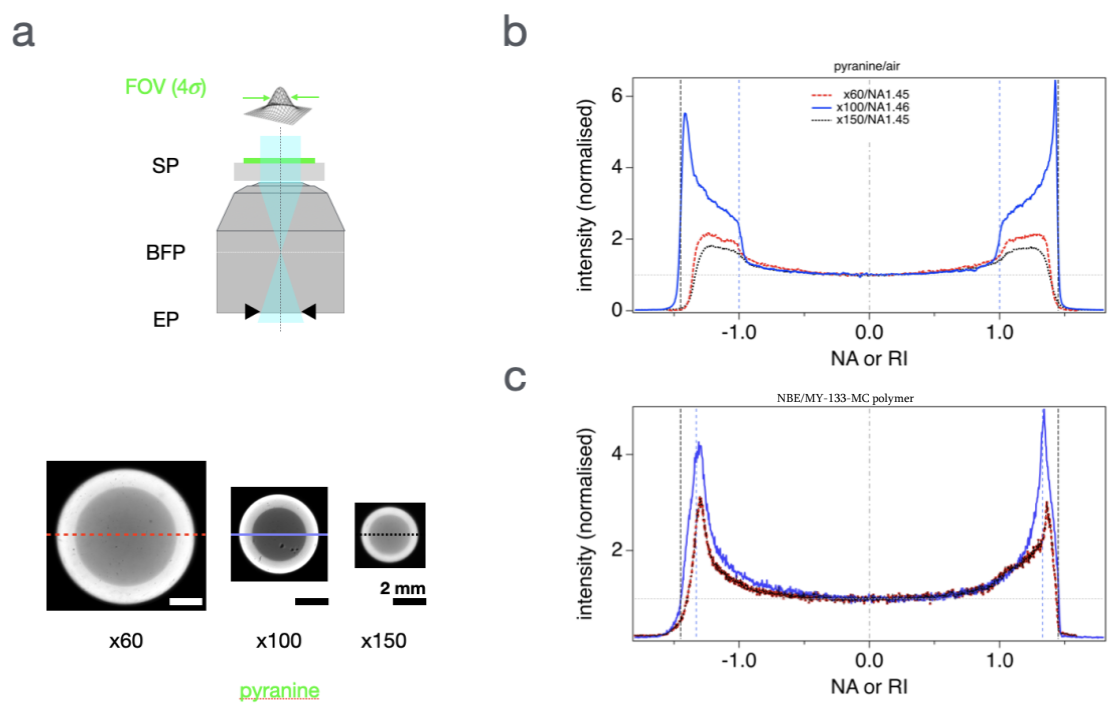

FIGURE 9. *Intensities and BFP-derived radiation patterns are objective-dependent*. (a), *top*, experimental paradigm. The entrance pupil (EP) of the objective lens and the focusing angle of the incoming beam determine the nominal and illuminated field-of-view (FOV), respectively, - measured as the $4\sigma$-diameter from a 2-D Gaussian fit with the SP image data (see fig.S17); SP - sample plane; BFP - back-focal plane. *Bottom*, example BFP images of a µm-thin pyranine dye film on BK-7 glass, for (from *left* to *right*) the PlanApo ×60/1.45 Oil TIRF, αPlan-Apochromat ×100/1.46 Oil DIC M27 and UAPON150X0TIRF ×150/1.45 objective, respectively. Cross-sections indicate regions used for measuring intensity profiles and color-code in (b). To better compare the profiles, fluorescence was normalized to the on-axis intensity and the pupil radii expressed as NAs rather than mm. Note the higher intensity and more peaked radiation pattern at higher NAs for the ZEISS lens. (c), same, for a bright, homogenous 8-nm thin nano-bead emitter (NBE) layer overlayed with a transparent, non-fluorescent polymer (MY-133-MC, $n = 1.33$). Albeit less pronounced, the ×60 and ×150 lenses performed similarly, and again a broader and better capture of high-angle information was observed for the ×100 lens. Color code as before.

Perhaps rather than reflecting only optical differences, the more efficient and higher-angle fluorescence capture seen with the ZEISS objective partly results from properties of the emission pattern of the dye film itself? For example, the high-intensity ring could be indicative of the presence of crystalline pyranine and mark the critical angle of another RI fluorophore environment. We tested this hypothesis by using a thinner, more controlled test sample instead. We prepared homogenous, bright, 8-nm thin fluorescent layers from a new class of bright, homogeneous green-fluorescent (ex./em. = 501/535 nm) nano-bead emitters (NBEs, (43, 44)) that were covered with a protective layer of transparent non-fluorescent My-133-MC polymer, having a RI of 1.33 (see ref. (31) for details). Conducting the same type of experiment and analysis as before, we recorded BFP images (not shown) and extracted intensity cross-sections, Fig. 9c. The calculated SAF/UAF ratios were close for the two Olympus objectives ($R = 0.14$ and $0.16$ for the ×60 and ×150 objective, respectively) but again double for the ZEISS lens, $R = 0.35$. Thus, with its higher NA and better transmission, the ×100/NA1.46 objective consistently outperformed the ×60 and ×150 lenses irrespective of the fluorophore layer. Of note, even the tiny difference in the $NA_{eff}$ between the ×60 and ×150 objectives (1.427 vs. 1.431, respectively) translates into significantly different SAF/UAF ratios, underpinning the importance of selecting and testing objectives for the highest possible NA, prior to buying these relative expensive objective lenses.

### What is the 'right' pixel size for BFP images?

Using objectives with similar NA does not (or very little) affect the optimal spatial sampling rate for SP images. In fact, for all three used objectives, a 4*f*-relay telescope is required to provide the extra magnification to match the pixel size in the SP to half the objectives' point-spread function (PSF), Fig. 10a. Knowing the objective's NA and the detector pixel size, we calculated the minimal total magnification $MM'$ needed to fulfil Nyquist's sampling theorem, Table 2. However, while it is straightforward to know the optimal pixel size for SP images, it is not so clear what is the correct sampling rate for BFP images.

The BFP or *k*-space contains the angular-spectrum representation, or spatial-frequency information, of the imaged object. Intimately related to its momentum ($h\mathbf{k}$), the BFP image can be viewed as the in-plane momentum ($k_x$,$k_y$) distribution of the imaged light. In the focal plane (SP), the lateral width of the point-spread function (PSF) is $\sim\lambda/(2NA)$ over a given FOV, which makes a number of modes $\sim$(FOV/PSF)$^2$ and the required pixel size $\Delta x = \Delta y = \Delta w \sim$ FOV/2PSF. In the BFP, the Fourier representation of the same signal contains the same number of spatial modes with an extent of the spatial frequency support equal to 2NA/$\lambda$ and a 'relevant' frequency sampling range $\sim$1/FOV, Fig.

. Thus, the larger the FOV (i.e., the lower the magnification of the objective lens), the more pixelated the BFP must be.

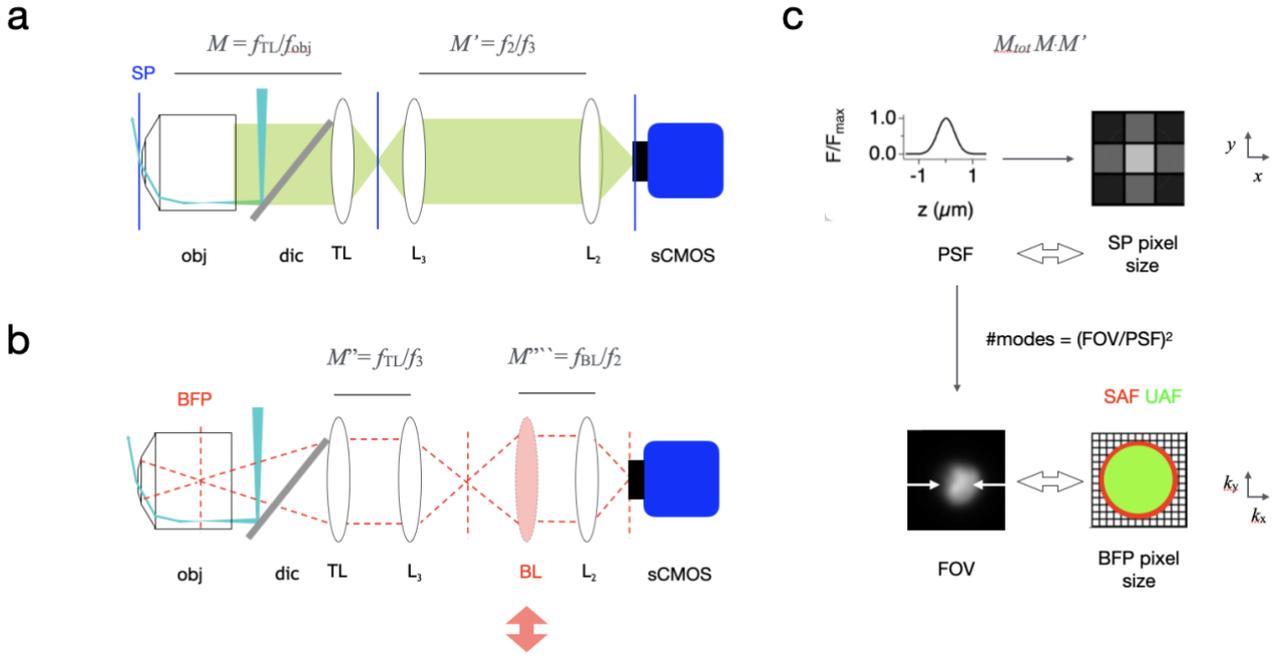

FIGURE 10. *Sampling SP and BFP images and pixel-size considerations.* (a), schematic layout of the emission optical path, consisting of a microscope and 4-*f* relay telescope. Total magnification is $MM'$ Abbreviations: SP - sample plane, obj - objective lens, dic - dichroic, TL - tube lens, L - lens, sCMOS - scientific complementary oxide sensor camera. *Blue*, conjugate field planes (SPs). *Turquoise* and *green* rays indicate TIRF excitation and imaged fluorescence, respectively. (b), same, with the Bertrand lens (BL, *red*) in place, which shifts the imaged plane to the objective's BFP. *Red dashed* lines identify conjugate aperture planes (equivalent BFPs). *Red dashed* lines are peripheral rays. Total magnification of the phase telescope is $M'''M'''$. (c), *top*, in the SP, the total magni-fication required is set by the NA of the objective lens, the physical size of the detector pixels and and the requirement to fulfil Nyquist's sampling theorem. The effective pixel size in the SP must not exceed half of the width of the point spread function (PSF) $\approx \lambda/(2NA)$ . *Bottom,* in the BFP, the field-of-view (FOV) is the relevant measure. As the same information transits the SP and BFP, the number of modes required to sample the SP $\approx$(FOV/PSF)$^2$ is required to preserve the information content of the BFP image, too.

We assume the high-NA lenses used here to obey Abbe's sine condition, i.e., the BFP radius $r_\theta$ and polar beam angle $\theta$ are related by $r_\theta = f_{obj} \cdot n_2 \cdot \sin\theta = f_{obj} \cdot NA$. These objectives can be considered an ideal Fourier transformer. For such objectives, we can assume the $k$-space resolution to be uniform over the entire $k$-space. We can hence turn to information theory: Since an (ideal) objective lens preserves information, the number of 'resolvable points' (Airy discs) in the SP and BFP must be equal. Therefore, if the FOV measures, say, 100 Airy discs across, the BFP should be sampled with twice as many pixels to fulfil the Nyquist theorem, Fig. 10c. For the three high-NA objectives used in this study, the required pixel sizes and magnifications are compiled in Table 2, and they allow us to draw some interesting conclusions. (*i*), as the objective magnification $M$ does not appear in the total magnification $M''\cdot M'''$ of the phase telescope, the effective pixel size of BFP images is the same for all objectives. Thus, to account for the different FOVs, camera binning[5] must be used to fine-tune the BFP pixel size for any given objective if we want to avoid oversampling; (*ii*), SP and BFP images generally will require different binning; (*iii*), the lower the magnification $M$ of the objective (i.e., the larger the FOV) the more pixels are required to sample the BFP as $\Delta k \propto 1/\text{FOV}$.

<u>Table 2</u>

*Optimal sample- and back-focal plane sampling.*

| Objective | SP | | | | | | BFP | | | |
|---|---|---|---|---|---|---|---|---|---|---|
| *Supplier M, NA* ($M_{eff}$, $NA_{eff}$) [a] | *FN* | *PSF* [b] | *FOV* [c] *(FOV_{eff})* [d] | *#px_{SP}* [e] | *px size, calculated (measured)* [g] | *SP bin* [h] | *EP* [j] (mm) | *#px_{BFP}* [k] | *px size, calculated (measured)* [m] | *BFP bin* [n] |
| | (mm) | (nm) | (μm) | | (nm) | | | (px) | (μm) | |
| OLYMPUS ×60/1.45 (×55/1.427) | 22 | 235 | 452 (50) | 3,846 (426) | 42 (58) | ×2 | 8.6 | 1,632 | 5.04 (5.2) | ×1 |
| ZEISS ×100/1.46 (×100/1.465) | 25 | 229 | 250 (34) | 2,183 (297) | 23 (33) | ×2 (×4) [i] | 4.83 | 940 | 5.04 (5.2) | ×2 |
| OLYMPUS ×150/1.45 (×137.5/1.431) | 22 | 235 | 181 (22) | 1,540 (187) | 17 (24) | ×4 | 3.43 | 652 | 5.04 (5.2) | ×2 |

*Abbreviations*:

   *M* - magnification, *NA* - numerical aperture, *FN* - field number, *PSF* - (width of the) point-spread function, *FOV* - field-of-view, *px* - pixel, *EP* - entrance pupil.

*Footnotes:*

[a] $M_{eff} = f_{TL}/f_{obj}$ but with the ZEISS tube lens used here ($f_{TL}$ = 165 mm); $NA_{eff}$ measured as in <u>ref. (26)</u>

[b] calculated Rayleigh resolution, $\approx 0.61\lambda/\text{NA}_{eff}$, at $\lambda$ = 550 nm

[c] calculated FOV, i.e., field number (FN) divided by $M_{eff}$, (measured) - this is the FOV, e.g., in the eyepieces

[d] measured, effectively illuminated FOV, which depends on the tightness of focussing of the excitation beam in the BFP, measured from fitting an ellipse with the image, taking the 4$\sigma$-diameter of a 2-axes Gaussian fit, and defining the diameter of an equivalent circle as $FOV_{eff}$.

[e] calculated as $2FOV/PSF$, this is the number of resolution elements needed across the FOV limited by the FN; the value in brackets is $2FOV_{eff}/PSF$, the number of pixels across the lit FOV.

[f] calculated as 6.5 μm/(2.8·$M_{eff}$), i.e., the detector pixel size divided by the total magnification of the microscope

[g] measured upon bright-field illumination, using a μm ruler (Thorlabs R1L3S2P), without binning

[h] possible binning to still be diffraction-limited, i.e., closed even integer to $PSF$/(2 measured *px size*)

[i] 'allowed' binning would be ×3.47

[j] Entrance pupil diameter (EP) at the back aperture of the objective, calculated as EP = 2 $NA_{eff}$ $f_{obj}$

[k] measured $EP_{eff}$ in pixels from intensity cross sections of BFP images

[l] calculated effective pixel size = (6.5 μm pixel size on the sCMOS sensor divided by the total magnification of the phase telescope (x3).

[l] measured upon bright-field illumination, using a mm ruler (Pepler Optics), without binning

[m] calculated as #px(SP)/ the number of points needed to sample the $FOV_{eff}$.

[n] calculated as total even number smaller than the total pixel number pixel (2048) divided by #px_{BFP}.

However, it is a bit artificial to relate the sampling frequency in the BFP to the image FOV. If the BFP is sampled to measure aberrations, for instance, the relevant frequency range should rather be chosen according to the spectral content of aberrations. In the case of SAF detection, however, the FOV *is* the relevant parameter to be considered because it will determine the resolution (the sharpness) of the SAF ring (in the same manner as the sharpness or size of the focused laser spot in the BFP determines the lit FOV in objective-type TIRF - again, here SAF and TIRF behave as optical reciprocals). Thus, for the ×60, ×100 and ×150 objectives, we prescribe BFP image pixel sizes of at least 8 μm, 14 μm and 20 μm, respectively. Our GUI automatically takes into account these differences when the Bertrand lens (BL) is inserted into the collection optical path. A corollary is that the small pixels of the used sCMOS detector (6.5 μm) permit on-chip binning and increase sensitivity compared to the typically bigger pixels of EMCCD cameras. However, unlike for (EM)CCD cameras, binning does not change the frame rate for sCMOS chips, as binning is performed in the FPGA *after* the image is read out from the sensor.

---

[5] Binning is the procedure of combining the signal from a number of adjacent pixels into an output for a single pixel (super-pixel). For 2×2 binning, an array of 4 pixels becomes a single larger super-pixel, reducing the overall number of pixels that need to be readout and also reducing the resolution available.

**DISCUSSION**

Our study explores how fluorophore distributions and experimental parameters affect the  effective fluorescence emission patterns measured on BFP images. We also present a comprehensive work-flow for BFP image analysis and quality assessment. The main findings of our work are:

(*i*), as expected, a homogeneous dye solution yields near-uniform emission patterns dominated by undercritical (UAF) emission, while a thin film of the same fluorophore deposited on the interface produces highly directional, supercritical angle fluorescence (SAF), with strong emission beyond the critical angle (NA > 1.33). Thus, the typically highly corrected high-NA lenses used for objective-type TIRF can be assumed to obey Abbe's sine condition and can be considered as near-ideal Fourier transformers.

(*ii*), in a real-world experiment, image imperfections like asymmetry, non-circularity, specs, etc. can arise from various sources (sample tilt, dust, non-uniform dye distribution) but the here introduced metrics including ovality (*OV*), circularity (*CI*), and concentricity offset (*CO*) can reliably assess BFP image quality and is a first, important step to reproducible BFP imaging;

(*iii*), automated dual-thresholding methods segment BFP images into SAF, UAF, and background zones work best for near-interface fluorophores or TIRF-excited samples, and less so for thick and and uniform fluorophore distributions or EPI excitation. This is good news for combinations of TIRF excitation and SAF detection, less so for EPI or confocal geometries resulting in appreciable volume excitation;

(*iv*), optimal SP- and BFP-image pixel size scale inversely with the objective's NA and field of view (FOV), respectively. Adaptive on-chip binning must be used when toggling between SP- and BFP image acquisitions to improve signal-to-noise ratio, which is critical for dim samples or short exposure times. Binning does not compromise linearity or the calculated SAF/UAF intensity ratio;

(*v*), we develop a sampling theory for SAF-based techniques: as BFP images represent the angular (*k*-space) distribution of fluorescence, proper BFP image pixel size follows Nyquist sampling principles based on the number of resolvable modes in the effectively lit FOV. One consequence is that lower magnifications require finer sampling in the BFP.

(*vi*), the use of a flipable Bertrand lens (BL) allows switching between SP and BFP imaging. BFP images depend on BL defocus, making BL alignment critical, particularly when using multiple objectives. One the other hand, BFP images are much less sensitive to sample defocus than SP images. Parameters like the SAF/UAF ratio remain stable over a broader *z*-range, making BFP imaging ideal for moving objects or long-term imaging. Also, a BFP-based rapid course focussing of the sample stage or objective focus drive could be an interesting strategy before switching to a finer (piezo) SP-based fine focus.

We provide a LABVIEW GUI for multi-angle TIRF/epifluorescence excitation and SP and combined SP and BFP imaging. No such software is currently available. Instead, common image acquisition software requires to adapt both exposure time, binning and region-on-chip each time when changing between SP- and BFP-image acquisitions, making rapid alternate SP and BFP imaging impossible. Also, no public software supports nested variable azimuthal or polar-incidence angle acquisition with time-series, BL toggling or spectral imaging. Our GUI treats and annotates these different images separately and it allows for complex, nested multi-parameter acquisitions. It also sets up meaningful file names sorts files into consistent pathways, which considerably facilitates data access and handling. Our GUI is available upon request.

**Two evanescent waves are better than one: 'coplanar' surface imaging**

Combining Total Internal Reflection Fluorescence (TIRF) with Supercritical Angle Fluorescence (SAF) (33) is attractive for advanced biological microscopy because it gives control over both the distribution of excitation light and the directional emission -which leads to a better background rejection and dramatically improves axial resolution, contrast, and sensitivity near the cell membrane. TIRF restricts the excitation light to a thin (~100–200 nm) region just above the glass–sample interface via an evanescent wave, but in objective-type TIRF and in a realistic biological experiment the 'true' penetration depth is often unknown (see ref. (35) and references therein for a recent discussion). SAF controls emission: It selectively detects only the portion of emitted fluorescence that exits the sample at supercritical angles, which only occurs when fluorophores are very close to the interface. Together, TIRF and SAF act like a double surface filter - one on the way in, one on the way out. Whereas we used here essentially BFP imaging and studied the retrieval of information from the fluorophore radiation pattern, Ries and Léveque-Fort proposed a technique called 'virtual' SAF (*v*SAF) (18, 33, 36) which uses computational post-processing to allow selective SAF detection during wide field SP imaging while maintaining a clear detection NA. Instead of physically filtering SAF with a ring aperture, it emulates SAF detection on a standard high-NA widefield microscope and applies full-aperture and UAF image subtraction to extract the SAF signal. The potential of combining TIRF and SAF is seen in single-molecule localization microscopies (SMLMs) where SAF detection is complementary to PSF-engineering and aperture-filtering strategies. Fluorophore height maps are calculated from image ratios. However, *v*SAF assumes a constant and symmetric BFP image and enforces aperture filtering by an iris stepping down the detection NA to $n_1$ to generate a 'UAF-only' image. Thus, while imaging the SP, *v*SAF suffers from identical limitations and depends on BFP image properties in the same manner as we discussed here.

Directly imaging the BFP instead using aperture filtering enhances the information content of near-interface microscopies: the radiation pattern has been used for microrefractometry (23, 24), for imaging the wetting as well as imperfections of nm-thin films (25), for the calibration of nanometric axial rulers that can serve as calibration tools for axial super-resolution microscopies (31), and for 3-D analyses of TIR-FCS data (12, 49). The framework developed in this paper paves the way for detecting time-dependent changes in these parameters and allows - with appropriate fluorophores chemically or genetically addressed to specific sub-cellular locations or organelles - for imaging sub-cellular dynamics at unprecedented detail (see the companion paper, Pt.II).

Considering the sensitivity of both TIRF and SAF on alignment, acquisition parameters and sample positioning, and given the wealth of information and degree of quality control that can be accessed through BFP image analysis, it is surprising how little use has been made of combined SP and BFP imaging. Liu *et al.* (2020) analyzed the excitation-light distribution in the BFP for the alignment of spinning-TIRF and obtain a more homogeneous illumination (30). In a similar application, BFP-image analysis was used for (polar) incidence angle calibration (28-30). Other studies investigated, respectively, the effects of focal drift (34) and how different fitting algorithms affect the precision of BFP-image segmentation (24) but, overall, the benefits of direct BFP imaging seem largely underrated.

## Conclusion

Our study makes a strong case for BFP image analysis in the context of near-interface fluorescence imaging. BFP image analysis allows not only to uncover additional fluorophore parameters like orientation, surface proximity, and local RI but it also facilitates the interpretation of TIRF, SALM and *v*SAF sample-plane images. Beyond its relevance for nm-axial metrology, our work clearly illustrates the potential of combined TIRF excitation and SAF/UAF ratio analysis for biological super-resolution and single-molecule localization imaging. Such applications will benefit from the seamless integration of BFP and SP-image acquisition as offered by our dedicated GUI. We hope that the systematic and rigorous investigation of how different experiment parameters affect BFP images will inspire researchers and industrials to make use of this additional information that can be obtained with comparably little effort on a standard (inverted) microscope.

**Acknowledgements.**

We thank Drs. Yossi Abulafia (Bar-Ilan University), Dan Axelrod (U Michigan), Baptiste Blochet, Marc Guillon (both SPPIN), and Alexander Jesacher (U Innsbruck) for helpful discussions. Patrice Jegouzo and Valentin Pierrat (*plateforme de prototypage*, CNRS UMS 2009, INSERM US 36, BioMedTech Facilities) are acknowledged for custom fine mechanics.

**Funding.**

Supported by the European Union (H2020 Eureka! Eurostars, 'NANOSCALE', (E!12848), https://nanoscale.sppin.fr, to MO and AS), and a PHC Maimonide collaboration grant (MAE/MESRI) to MO and AS. AS acknowledges support from the French Ministry of Foreign Affairs as well as the French Embassy in Tel Aviv (Chateaubriand). We are grateful for mobility support from a CNRS-LIA, 'ImagiNano', and a CNRS-IRP 'OMNI-tools', funding from the CNRS *mission internationale* and FranceBioImaging (a large-scale national infrastructure initiative, FBI, ANR-10-INSB-04, Investments for the future). OS received funding from a Université Paris Cité pre-maturation grant. OS acknowledges support from University Paris Cité (VAE) and SPPIN.

**Disclosures.**
The authors declare no conflicts of interest.

**Data availability statement.**
Data can be made available upon reasonable request.